\documentclass[a4paper,fleqn]{cas-dc}

\usepackage[numbers]{natbib}
\usepackage{amsmath}
\usepackage{algorithm}
\usepackage{algpseudocode}
\usepackage{xcolor}
\def\tsc#1{\csdef{#1}{\textsc{\lowercase{#1}}\xspace}}
\tsc{WGM}
\tsc{QE}
\tsc{EP}
\tsc{PMS}
\tsc{BEC}
\tsc{DE}
\newcommand{\mathcolorbox}[2]{\colorbox{#1}{$\displaystyle #2$}}
\definecolor{mypink}{RGB}{254,194,204}
\definecolor{Gray}{gray}{0.9}
\def \name{\textsc{AlterMILP}\xspace}

\usepackage{soul}
\usepackage{xcolor}
\usepackage{tcolorbox}
\usepackage{pifont}
\usepackage{subfigure}

\algrenewcommand{\algorithmiccomment}[1]{\textbf{//}\,#1}
\begin{document}
\newcommand{\yy}[1]{\textcolor{red}{\bf\small [#1 --YY]}}
\newcommand{\sy}[1]{\textcolor{blue}{\bf\small [#1 --SF]}}
\newcommand{\jk}[1]{\textcolor{orange}{\bf\small [#1 --JK]}}
\shorttitle{Alternative MILP Optimization for Job Scheduling and Data Allocation in Grid Computing}

\title [mode = title]{Alternative Mixed Integer Linear Programming Optimization for Joint Job Scheduling and Data Allocation in Grid Computing }                      



\author[1]{Shengyu Feng}
\cormark[1]
\fnmark[1]
\ead{shengyuf@cs.cmu.edu}

\author[1,2]{Jaehyung Kim}
\fnmark[1,2]
\ead{jaehyungk@yonsei.ac.kr}           
                
\author[1]{Yiming Yang}
\ead{yiming@cs.cmu.edu}

\author[3]{Joseph Boudreau}
\ead{boudreau@pitt.edu}

\author[4]{Tasnuva Chowdhury}
\ead{tchowdhur@bnl.gov}

\author[4]{Adolfy Hoisie}
\ead{ahoisie@bnl.gov}

\author[3]{Raees Khan}
\ead{raees.ahmad.khan@cern.ch}

\author[4]{Ozgur O. Kilic}
\ead{okilic@bnl.gov}

\author[5]{Scott Klasky}
\ead{klasky@ornl.gov}

\author[3]{Tatiana Korchuganova}
\ead{TAK245@pitt.edu}

\author[4]{Paul Nilsson}
\ead{Paul.Nilsson@cern.ch}

\author[6]{Verena Ingrid Martinez Outschoorn}
\ead{Verena.Martinez@cern.ch}

\author[4]{David K. Park}
\ead{dpark1@bnl.gov}

\author[5]{Norbert Podhorszki}
\ead{pnorbert@ornl.gov}

\author[4]{Yihui Ren}
\ead{yren@bnl.gov}

\author[4]{Frederic Suter}
\ead{suterf@ornl.gov}

\author[4]{Sairam Sri Vatsavai}
\ead{ssrivatsa@bnl.gov}

\author[7]{Wei Yang}
\ead{yangw@slac.stanford.edu}

\author[4]{Shinjae Yoo}
\ead{sjyoo@bnl.gov}

\author[4]{Tadashi Maeno}
\ead{tadashi.maeno@cern.ch}

\author[4]{Alexei Klimentov}
\ead{aak@bnl.gov}

\affiliation[1]{organization={Carnegie Mellon University},
                city={Pittsburgh},
                state={PA},
                country={USA}}
\affiliation[2]{organization={Yonsei University},
                city={Seoul},
                country={South Korea}}
\affiliation[3]{
organization={University of Pittsburgh},
                city={Pittsburgh},
                state={PA},
                country={USA}}
\affiliation[4]{
organization={Brookhaven National Laboratory},
                city={Upton},
                state={NY},
                country={USA}}
\affiliation[5]{
organization={Oak Ridge National Laboratory},
                city={Oak Ridge},
                state={TN},
                country={USA}}
\affiliation[6]{
organization={University of Massachusetts, Amherst},
                city={Amherst},
                state={MA},
                country={USA}}
\affiliation[7]{
organization={SLAC National Accelerator Laboratory},
                city={Menlo Park},
                state={CA},
                country={USA}}     
\cortext[cor1]{Corresponding author}
\fntext[1]{Equal contribution}
\fntext[2]{Work done at Carnegie Mellon University}

\begin{abstract}
This paper presents a novel approach to the joint optimization of job scheduling and data allocation in grid computing environments. We formulate this joint optimization problem as a mixed integer quadratically constrained program. To tackle the nonlinearity in the constraint, we alternatively fix a subset of decision variables and optimize the remaining ones via Mixed Integer Linear Programming (MILP). We solve the MILP problem at each iteration via an off-the-shelf MILP solver. Our experimental results show that our method significantly outperforms existing heuristic methods, employing either independent optimization or joint optimization strategies. We have also verified the generalization ability of our method over grid environments with various sizes and its high robustness to the algorithm hyper-parameters.
\end{abstract}

\begin{keywords}
Job scheduling \sep Date allocation  \sep Mixed integer linear programming \sep Grid computing environments \sep High performance computing
\vspace{2in}
\end{keywords}

\maketitle

\section{Introduction}
Grid computing has emerged as a powerful tool for processing the data-intensive jobs in modern scientific research, such as the particle physics \citep{Evans2009TheLH}, biology \citep{Au2009Grid}, astronomy \citep{Benacchio07Grid}, and earth science \citep{Renard09Grid}. 
Its distributed framework allows for the efficient integration and utilization of diverse resources in the environment, which could be roughly classified into two primary types, the computational resources and storage resources. 
How to properly coordinate these resources has largely decided the efficiency of the grid computing, e.g., the total processing time (makespan), the system throughput or the resource utilization, and it has been a central focus of the recent research in high-performance computing \citep{Tyagi2018Survey, HUSSAIN2013709}.

The coordination of resources can be further dissected into two critical aspects: job scheduling and data allocation. 
Job scheduling involves assigning jobs to computational nodes and determining their priority within the queue of each node, a process that is vital for the effective parallel execution of distributed jobs. 
Each computational node is a basic computing unit in the grid environment, subject to a limited memory. 
When the assigned jobs exceed its memory, it will put the jobs with the lower priority in its local waiting queue, leading to a running delay.
Therefore, job scheduling needs to balance the number of jobs assigned to each computational node in order to efficiently leverage the parallelism of the distributed computing. 
On the other hand, the running time of each job consists of both the data transmission delay and the execution time. 
Due to the different availability of each data object for downloading, it is also important to manage the priority for the jobs within the same computational node to let the job be executed first if its required data objects are ready. 
Conversely, data allocation (or data replication) focuses on selecting appropriate storage nodes to cache data, thereby minimizing data transmission delays when the computational node retrieves input data objects for each job.
As the bandwidths between computational nodes and storage nodes differ, the assignment must take into account the required objects of the jobs at each computational node. 
Both job scheduling and data allocation are recognized as NP-hard problems \citep{Du1989Complexity, Shmoys1993Approximation}, and numerous efficient heuristics have been developed to address these challenges approximately in the last few decades \citep{BRAUN2001810, Xhafa2010Computational, Mishra2014Survey, AMJAD2012337, Govardhan2024survey}. 
Although significant progress has been made in optimizing each problem independently, simply merging their optimization strategies does not result in good overall efficiency due to the intrinsic interconnection of these two problems.
For instance, data allocation heuristics must account for the distribution of input objects across computational nodes, while the job schedule should be optimized based on data object availability. 
Consequently, jointly optimizing both aspects remains a substantial challenge in real-world applications, necessitating innovative approaches to effectively enhance performance results.

\begin{figure*}[t!]
    \centering
    \includegraphics[width=0.9\linewidth]{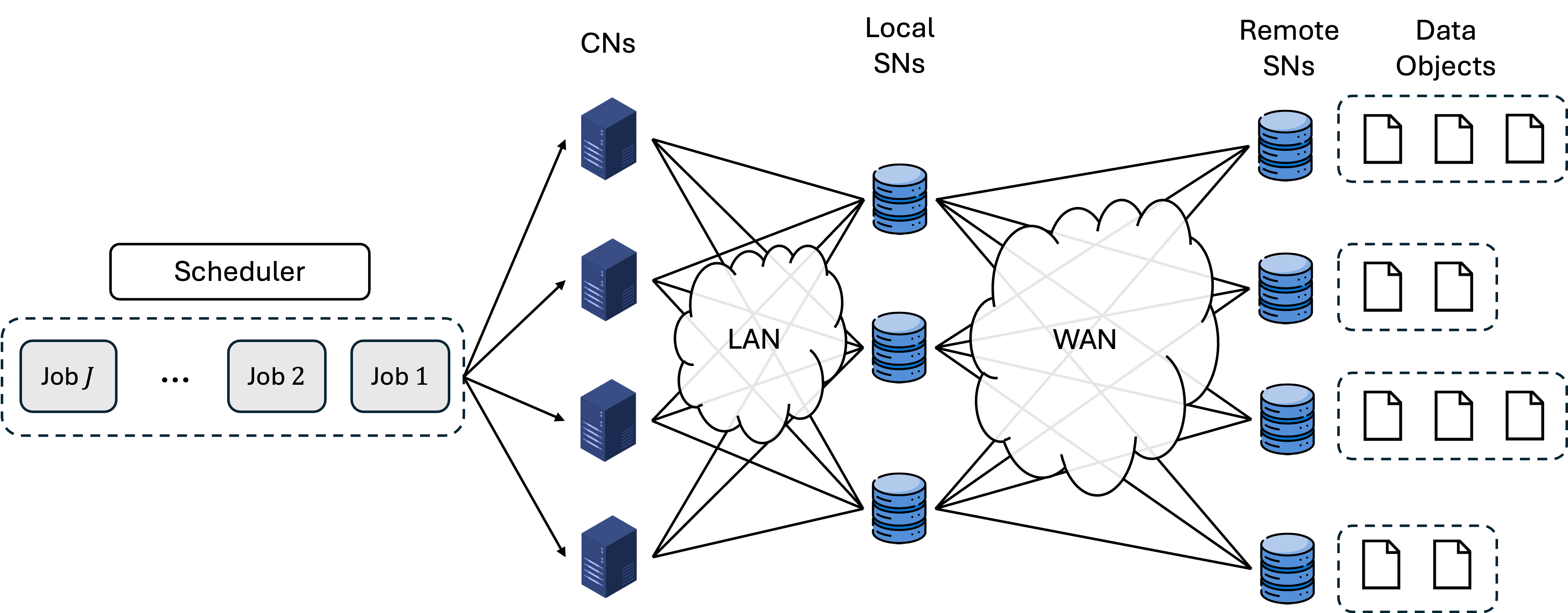}
    \caption{Overview of the grid computing environment. It consists of the jobs, computational nodes (CNs), storage nodes (SNs) and data objects. The remote SNs are connected with local SNs through a low-speed Wide Area Network (WAN) and CNs are connected with local SNs via a high-speed Local Area Network (LAN).}
    \label{fig:framework}
\end{figure*}

The interaction between job scheduling and data allocation makes it hard to design an efficient heuristic catering to the structure of the joint optimization problem. 
Therefore, a common approach to tackle this joint optimization problem is through the meta-heuristic, which offers a unified framework for the combinatorial optimization problem independent of the problem structure. 
Some common meta-heuristics used to address this challenge include the genetic algorithm \citep{Phan05Co}, artificial bee colony optimization \citep{TAHERI20131564} and particle warm optimization \citep{liu2013swarm}. 
Due to recent advances in deep learning, there has also been an increasing popularity in solving this type of joint optimization problem through deep reinforcement learning \citep{wei2021joint1, wei2021joint2, ZENG2024121}. 
Although promising, these general frameworks typically fall short in utilizing the problem structure, and thus unavoidably end up with a suboptimal performance. 
To overcome such an issue, we first formulate the joint optimization problem as a mixed integer quadratically constrained program and identify the nonlinear constraints that prevent the efficient utilization of problem structure.
Inspired by coordinate descent algorithms \citep{Wright2015CoordinateDA}, we introduce an iterative optimization approach called Alternative Mixed Integer Linear Programming (\name{}), which alternately fixes a subset of decision variables and optimizes the remaining ones using mixed integer linear programming (MILP). Compared to the original nonlinear program, the resulting mixed integer linear program is significantly easier and more efficient to solve.
We then use an off-the-shelf MILP solver, Gurobi \citep{gurobi}, to address the sub-problem at each iteration and gradually improve our solution to the joint optimization problem. The works most closely related to ours are Ko el al. \citep{Ko2019MIQP} and Govardhan et al. \citep{TAHERI20131885}. 
Ko el al. \citep{Ko2019MIQP} also employ the mixed integer programming (MIP) technique in joint optimization, but directly tackling the quadratic constraints via mixed integer quadratic programming (MIQP). 
It thus suffers from a bad scalability and we address this issue by decomposing the quadratic constraints into linear ones. 
Govardhan et al. \citep{TAHERI20131885} decompose the joint optimization problem into job scheduling and data allocation and alternatively updates each problem. 
However, their way of decomposition does not really get rid of the nonlinear constraint and they optimize each problem via the Hopfield neural network algorithm, which we find empirically less effective than MILP in finding a high-quality solution within the same computational budget (Section~\ref{sec:4.2}).
In a nutshell, \name{} has assimilated the strength of MILP and the coordinate descent strategy, representing the first research effort to apply MILP in solving the joint optimization of job scheduling and data allocation in grid computing environments.
 

Our empirical results have shown that our \name{} method significantly outperforms the previous heuristic-based methods, either independent optimization or joint optimization methods in minimizing the makespan of grid computing. 
\name{} demonstrates a consistent advantage in grid computing environment with various sizes, being efficient in the decision time and robust to the algorithm hyper-parameters. 
All of this empirical evidence suggests the promising potential of \name{} to efficiently coordinate heterogeneous resources in grid computing.

\section{Problem Statement}
\begin{table}[h!]
\caption{List of notations.}\label{tab:notation}
\begin{tabular*}{\tblwidth}{@{}L|L@{}}
\toprule
Notation  & Definition \\ 
\midrule
$D$; $J$; $C$; $L$; $R$& \makecell[l]{The number of data objects, jobs, CNs, local\\ SNs and remote SNs}\\
\hline
$\mathcal{O}_j$ & The set of input data objects of job $j$\\
\hline
$\mathbf{s}$ & The size vector of data objects\\
\hline
$\mathbf{p}$ & The processing speed vector of CNs\\
\hline
$\gamma$ & The computation coefficient \\
\hline
$\mathbf{u}$; $\mathbf{v}$; $\mathbf{e}$ &  \makecell[l]{The job start time vector, execution start\\ time vector and execution length vector}\\
\hline
$\mathbf{B}^{(1)}$; $\mathbf{B}^{(2)}$ & \makecell[l]{The bandwidth matrix between remote SNs \\ and local SNs, and the bandwidth matrix \\ between local SNs and CNs}\\
\hline
$\mathbf{t}$ & \makecell[l]{The remote transmission delay (from remote \\ SNs to local SNs) vector for data objects} \\
\hline
$\mathbf{X}$; $\mathbf{Y}$; $\mathbf{Z}$ & \makecell[l]{The job assignment matrix, job precedence \\ matrix and data assignment matrix}\\
\hline
$m$ & The  makespan \\
\bottomrule
\end{tabular*}
\end{table}

\subsection{Framework Overview}
We consider a grid computing environment as shown in Figure \ref{fig:framework}. 
The grid environment consists of four main elements: data objects, jobs, computational nodes, and storage nodes. 
We summarize all of our notations in Table \ref{tab:notation}.

\paragraph{Data objects.} The data object refers to any input file to be processed by the job. 
We assume that there are in total $D$ data objects and their sizes can be represented as a vector $\mathbf{s}\in\mathbb{R}^D_{>0}$, with $\mathbf{s}_d$ indicating the size of the data object $d$.
\paragraph{Jobs.} Each job is a computing program submitted by the user, which involves a series of computation operations on the input data objects, for example, $\mathcal{O}_j\subseteq\{1,\cdots,D\}$ for job $j$. 
For simplicity, we assume that the total amount of computation operations of a job is linearly proportional to the total size of its input objects, that is, $\gamma\sum_{d\in\mathcal{O}_j}\mathbf{s}_d$, where the computational coefficient $\gamma$ is a constant within the system. 
Instead of doing a first-in-first-out ordering, we consider a look-ahead mechanism where the jobs would not be dispatched until there are enough jobs held in the global scheduler, e.g., $J$ jobs.

\paragraph{Computational nodes.} The computational node (CN) is the basic computing unit in the grid environment, responsible for executing the computation operations of a job. 
Due to limited memory, each CN maintains a first-in-first-out local queue to process the submitted job sequentially. 
For simplicity, we assume that the CN can process only one job at a time in our framework. 
There are in total $C$ CNs in the environment, and we use a vector $\mathbf{p}\in\mathbb{R}^C_{>0}$ to represent the processing speed of all CNs. 
The execution time of the job $j$ in CN $c$ could be calculated as $\gamma\sum_{d\in\mathcal{O}_j}\mathbf{s}_d/\mathbf{p}_c$.

\paragraph{Storage nodes.} The storage node (SN) is used to store data objects and there are basically two types of SN: remote SNs and local SNs. 
All data objects are assumed to be initially stored in remote SNs, denoted as a vector $\mathbf{I}\in[1,R]^D$, and are connected to the computing resources via a low-speed Wide Area Network (WAN). 
Hence, downloading the data objects directly from the remote SNs would lead to a high data transmission delay. 
The local SNs are intended to reduce this latency by caching the previously used data objects, and the CNs are connected to local SNs via a high-speed Local Area Network (LAN). 
Instead of doing an on-demand download from the remote SNs, the local SNs replicate the data objects in parallel to the job running. 
The number of replications of each data object is usually restricted in practice, in order to satisfy the memory constraint of local SNs and avoid the replication inconsistency. 
In this work, we assume that each data object could only be replicated to one local SN.

We assume that there are $L$ local SNs and $R$ remote SNs. 
Let $\mathbf{B}^{(1)}_{rl}$ represent the bandwidth between remote SN $r$ and local SN $l$ and $\mathbf{B}^{(2)}_{lc}$ stand for the bandwidth between local SN $l$ and CN $c$. 
We assume that the transmission delay of the data object $d$ can be computed as $\mathbf{s}_d/\texttt{bandwidth}$. 
To avoid clutter in the notation, we simply use $td^{(1)}(d,l)$ to represent the remote transmission delay of the data object $d$ from the remote SN $r$ to the local SN $l$ and $td^{(2)}(d,l,c)$ to denote its transmission delay from the local SN $l$ to the CN $c$. 
In particular, the remote SN that hosts each data object is fixed in the system, so we omit it in $td^{(1)}$.
\newline

\begin{figure}[t!]
    \centering
    \includegraphics[width=\linewidth]{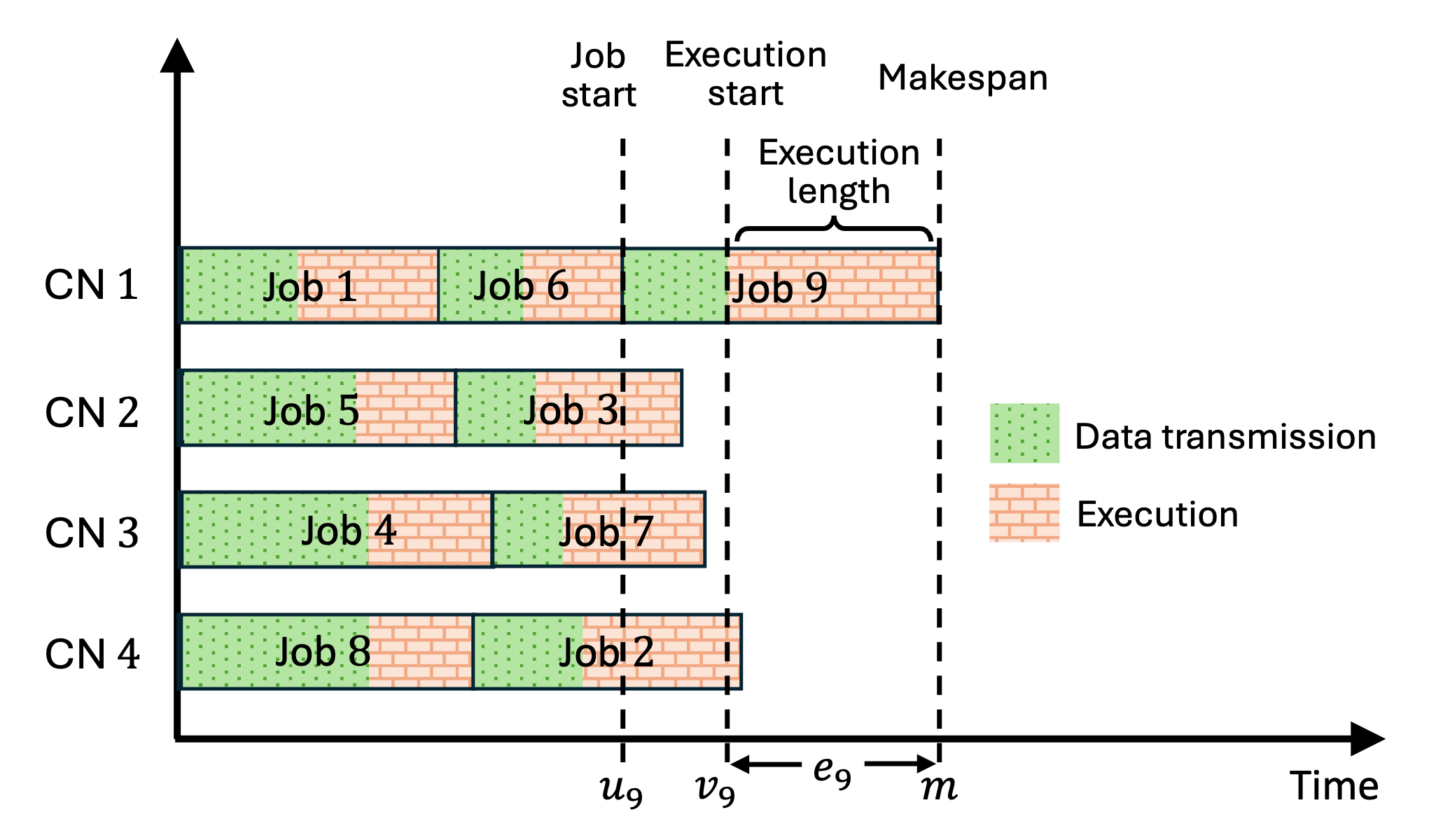}
    \caption{Illustration of the makespan. The running time of each job consists of both the data downloading time and execution time. The completion time of the last job marks the makespan of the job batch.}\label{fig:makespan}
\end{figure}

The whole pipeline of our framework can be summarized as follows. 
Users submit their jobs to a centralized scheduler, which begins the job scheduling and data allocation process once it has received $J$ jobs. 
The scheduler assigns each job to a CN and determines the priority of the job. 
If multiple jobs are assigned to the same CN, those with higher priority are processed first within the local queue. 
Concurrently, the scheduler allocates data objects to local SNs, and these objects are replicated in parallel during the job running. 
Upon the start of a job, the CN attempts to retrieve its input data objects from the local SNs. 
If an object transfer is still ongoing, the CN will wait for its completion before proceeding.
Execution of the job only begins once all input data objects are loaded into the CN's memory.
Ultimately, the makespan of the entire computational process is determined by the completion time of the last job. 
The actual running time of each job consists of both the data download time and the execution time. 
The whole process is illustrated in Pseudocode \ref{alg:process} and Figure \ref{fig:makespan}.

\begin{algorithm*}
      \caption{Pseudocode of the grid computing pipeline.}
      \begin{algorithmic}[1]
        \Procedure{Makespan}{$JobQueue$, $JobAssignment$, $DataAssignment$}
        \State \Comment{$JobQueue$: A queue of $J$ jobs, ordered in the descending priority}
        \State \Comment{$JobAssignment$: A $J$-dimensional vector of the assigned CN for each job}
        \State \Comment{$DataAssignment$: A $D$-dimensional vector of the assigned local SN for each data object}
        \State 
        \State \Comment{Initialize makespan, CN available time,  and data object available time} 
        \State $m=0$, $CT=\mathbf{0}^C$, $DT=\mathbf{0}^{D}$
        \State
        \State \Comment{Start the remote data transmission}
        \For{Data object $d=1,\cdots,D$}
        \State $DT[d]=td^{(1)}(d,DataAssignment[d])$
        \EndFor
        \State
        \State  \Comment{Start the job running}
        \For{Job $j$ in $JobQueue$}
        \State $\mathbf{u}[j]=CT[JobAssignment[j]]$ 
        \State 
        \State \Comment{Initialize the execution start time and the amount of computation operations}
        \State $\mathbf{v}[j]=\mathbf{u}[j]$, $operations=0$
        \State 
        \For{Data object $d\in\mathcal{O}_j$}
        \State $start=\max(\mathbf{u}[j], DT[d])$
        \State $\mathbf{v}[j]=\max(\mathbf{v}[j],  start + td^{(2)}(d,DataAssignment[d], JobAssignment[j]))$
        \State $operations = operations + \gamma*\mathbf{s}[d]$
        \EndFor
        \State
        \State \Comment{Start the job execution}
        \State $\mathbf{e}[j]=operations/\mathbf{p}[j]$
        \State $CT[JobAssignment[j]] = \mathbf{v}[j] + \mathbf{e}[j]$
        \State
        \State \Comment{Keep track of the largest job completion time}
        \State $m=\max(m, CT[JobAssignment[j]])$
        \EndFor
        \State Return $m$
        \EndProcedure
      \end{algorithmic}\label{alg:process}
\end{algorithm*}

\subsection{Problem Formulation}
We aim to minimize the makespan $m$ of $J$ jobs by jointly optimizing the job scheduling and data allocation in our framework. 
Here we use $\mathbf{u},\mathbf{v},\mathbf{e}\in\mathbb{R}_{\geq0}^J$, to denote the start time, the execution start time, and the execution length of all jobs, respectively. 
And $\mathbf{t}\in\mathbb{R}_{\geq 0}^D$ stands for remote transmission delays for all data objects. 
The makepsan is decided by the completion time of the last job, so it satisfies
\begin{equation}
    m\geq \mathbf{v}_j + \mathbf{e}_j, \quad\forall j\in[1,J].
\end{equation}

\paragraph{Job scheduling.} 
We further decompose job scheduling into two types of decision variables, job assignment and job ordering. 
The job assignment decides which CN to execute the job. 
We use a binary matrix $\mathbf{X}\in\{0,1\}^{J\times C}$ to represent the assignment of the job, where $\mathbf{X}_{jc}=1$ if the job $j$ is assigned to CN $c$. 
Since each job shall be assigned to exactly one CN, the assignment matrix should satisfy
\begin{equation}
    \sum_{c=1}^{C}\mathbf{X}_{jc}=1, \quad\forall j\in[1,J].
\end{equation}
The execution time of job $j$ could therefore be computed as 
\begin{equation}
    \mathbf{e}_j=\sum_{c=1}^C\mathbf{X}_{jc}(\gamma\sum_{d\in\mathcal{O}_j}\mathbf{s}_d/\mathbf{p}_c), \quad\forall j\in[1,J].
\end{equation}
The scheduler also decides the order or priority of each job. 
We use a precedence matrix $\mathbf{Y}\in\{0,1\}^{J\times J}$ to represent the order of the jobs, where $Y_{ij}=1$ means that the job $i$ has a higher priority than the job $j$, and $0$ otherwise. 
To make it a valid precedence matrix, we constrain $\mathbf{Y}$ to satisfy
\begin{align}
    &\mathbf{Y}_{ij}+\mathbf{Y}_{ji}=1, \quad\forall i,j\in[1,J], i\neq j; \\
    &\mathbf{Y}_{jj}=0, \quad\forall  j\in[1,J];\\
    \begin{split}
        &\mathbf{u}_j \geq A(\mathbf{Y}_{ij}(\mathbf{X}_{ic}+\mathbf{X}_{jc}-1)-1)+\mathbf{v}_i+\mathbf{e}_i,\\&\qquad\qquad\qquad\quad\forall i,j\in[1,J],i\neq j,c\in[1,C].
    \end{split}
\end{align}
where $A$ is a large constant such that the inequality always holds when $\mathbf{Y}_{ij}=0$ (job $i$ has a lower priority than job $j$) or $\mathbf{X}_{ic}+\mathbf{X}_{jc}\leq1$ (job $i$ and job $j$ are not assigned to CN $c$ at the same time).

\paragraph{Data allocation.}  The data allocation decides which local SN to replicate the data object. 
We also represent it with a binary assignment matrix $\mathbf{Z}\in\{0,1\}^{D\times L}$. 
Since each data object can only be replicated on one local SN, it should satisfy the constraint
\begin{equation}
        \sum_{l=1}^{L}\mathbf{Z}_{dl}=1, \quad\forall d\in[1,D].
\end{equation}
Besides, we can also express the remote transmission delay  of each data object as
\begin{equation}
        \mathbf{t}_d = \sum_{l=1}^L td^{(1)}(d,l)\mathbf{Z}_{dl}, \quad\forall d\in[1,D].
\end{equation}
Since the local data transmission does not start until the job starts and the input data objects get replicated on the local SNs, we obtain the constraints for the execution start time of each job as
\begin{align}
    \begin{split}
            &\mathbf{v}_j\geq  \mathbf{t}_d+ \sum_{l=1}^L\sum_{c=1}^Ctd^{(2)}(d,l,c)\mathbf{X}_{jc}\mathbf{Z}_{dl},\\&\qquad\qquad\qquad\qquad\qquad\quad\forall d\in\mathcal{O}_j, j\in[1,J]; 
    \end{split}\\
        \begin{split}
            &\mathbf{v}_j\geq \mathbf{u}_j + \sum_{l=1}^L\sum_{c=1}^Ctd^{(2)}(d,l,c)\mathbf{X}_{jc}\mathbf{Z}_{dl},\\&\qquad\qquad\qquad\qquad\qquad\quad\forall d\in\mathcal{O}_j, j\in[1,J]; 
    \end{split}
\end{align}
All the above constraints can be integrated into the following mixed integer quadratically constrained program (quadratic constraints highlighted in pink):

\begin{align}
    &\min_{m, \mathbf{X},\mathbf{Y},\mathbf{Z},\mathbf{u},\mathbf{v}, \mathbf{e}, \mathbf{t}} m \label{eq:11}\\
    s.t. \hspace{0.1cm} &   m\geq \mathbf{v}_j + \mathbf{e}_j, \quad \forall j\in[1,J];\\
    &\sum_{c=1}^{C}\mathbf{X}_{jc}=1, \quad \forall j\in[1,J];\\
    &     \mathbf{e}_j=\sum_{c=1}^C\mathbf{X}_{jc}(\gamma\sum_{d\in\mathcal{O}_j}\mathbf{s}_d/\mathbf{p}_c), \quad\forall j\in[1,J]; \\
    &\mathbf{Y}_{ij}+\mathbf{Y}_{ji}=1, \quad\forall i,j\in[1,J],i\neq j; \\
    &\mathbf{Y}_{jj}=0, \quad\forall  j\in[1,J];\\
    \begin{split}
    &\mathcolorbox{mypink}{\mathbf{u}_j \geq A(\mathbf{Y}_{ij}(\mathbf{X}_{ic}+\mathbf{X}_{jc}-1)-1)+\mathbf{v}_i+\mathbf{e}_i},\\
    &\qquad\qquad\quad\forall i,j\in[1,J],i\neq j, c\in[1,C];\label{eq:16}    
    \end{split}\\
    &\sum_{l=1}^{L}\mathbf{Z}_{dl}=1, \quad \forall d\in[1,D];\\
    & \mathbf{t}_d = \sum_{l=1}^L td^{(1)}(d,l)\mathbf{Z}_{dl} \quad \forall d\in[1,D];\\
    \begin{split}
            &\mathcolorbox{mypink}{\mathbf{v}_j\geq  \mathbf{t}_d+ \sum_{l=1}^L\sum_{c=1}^Ctd^{(2)}(d,l,c)\mathbf{X}_{jc}\mathbf{Z}_{dl}},\\&\qquad\qquad\qquad\qquad\quad\forall d\in\mathcal{O}_j, j\in[1,J]; \label{eq:20}
    \end{split}\\
        \begin{split}
            &\mathcolorbox{mypink}{\mathbf{v}_j\geq \mathbf{u}_j + \sum_{l=1}^L\sum_{c=1}^Ctd^{(2)}(d,l,c)\mathbf{X}_{jc}\mathbf{Z}_{dl}},\\&\qquad\qquad\qquad\qquad\quad\forall d\in\mathcal{O}_j, j\in[1,J]. \label{eq:21}
    \end{split}
\end{align}

\section{Joint Job Scheduling and Data Allocation via Alternative MILP Optimization}\label{sec:method}
In this section, we introduce our technique, called alternative MILP (\name{}), which efficiently addresses the formulated optimization problem for joint job scheduling and data allocation.

As highlighted in pink (Eqs. \ref{eq:16}, \ref{eq:20}, and \ref{eq:21}), the primary challenge in solving this problem stems from its quadratic constraints, which typically require mixed integer quadratic programming (MIQP) for direct resolution. 
However, MIQP methods are generally more complex and computationally intensive than mixed integer linear programming (MILP). 
Although existing MIP solvers are usually versatile enough for MIQP\footnote{\url{https://www.gurobi.com/resources}}, our empirical analysis reveals that their scalability in MIQP is significantly constrained (see Table \ref{tab:main_results} in Section \ref{sec:experiments}).
To address this challenge, we begin by identifying the unique structures inherent in the quadratic constraints of the joint optimization problem. 
We demonstrate that the problem can be efficiently decomposed into two distinct sub-problems, each characterized solely by linear and integer constraints, which is essentially mixed integer linear programs. 
The proposed \name{} method then alternates between solving each sub-problem using established MILP solvers, known for their efficiency. 
After a predetermined number of iterations, this approach yields the optimized variables. 
The detailed procedure of \name{} is outlined in Algorithm \ref{alg:altermilp}.  

\begin{algorithm}
      \caption{Pseudocode of {\name{}} algorithm.}
      \begin{algorithmic}[1]
        \Procedure{\name{}}{$T$}
        \State \Comment{$T$: Number of iterations}
        \State \Comment{Randomly initialize the job assignment matrix $\mathbf{X}$,  job precedence matrix $\mathbf{Y}$, and data assignment matrix $\mathbf{Z}$} 
        \State $\mathbf{X}^{(0)}, \mathbf{Y}^{(0)}, \mathbf{Z}^{(0)} \leftarrow \texttt{Random Initialization}$
        \State \Comment{Alternative optimization}
        \For{Iteration $t=1,\cdots,T$}
        \State $\mathbf{X}^{(t)}\leftarrow \arg\min_{\mathbf{X}}(\texttt{MILP}(\mathbf{X}|\mathbf{Y}=\mathbf{Y}^{(t-1)}, \mathbf{Z}=\mathbf{Z}^{(t-1)}))$
        \State $\mathbf{Y}^{(t)},\mathbf{Z}^{(t)} \leftarrow \arg\min_{\mathbf{Y},\mathbf{Z}} (\texttt{MILP}(\mathbf{Y},\mathbf{Z}|\mathbf{X}=\mathbf{X}^{(t)}))$
        \EndFor
        \State $\mathbf{X}^{*}, \mathbf{Y}^{*}, \mathbf{Z}^{*} \leftarrow \mathbf{X}^{(T)}, \mathbf{Y}^{(T)}, \mathbf{Z}^{(T)}$
        \State Return $\mathbf{X}^{*}, \mathbf{Y}^{*}, \mathbf{Z}^{*}$
        \EndProcedure
      \end{algorithmic}\label{alg:altermilp}
\end{algorithm}

\subsection{MILP Problem Decomposition by Selectively Fixing Decision Variables}\label{sec:3.1}

In our formulated problem for joint job scheduling and data allocation (Eqs. \ref{eq:11}-\ref{eq:21}), the quadratic nature arises from two distinct products of decision variables: $\mathbf{Y}_{ij}\mathbf{X}_{jc}$ (Eq. \ref{eq:16}) and $\mathbf{X}_{jc}\mathbf{Z}_{dl}$ (Eqs. \ref{eq:20} and \ref{eq:21}). 
These products share common decision variables, $\mathbf{X}$, related to the assignment of the job.
Our key strategy is to decompose the problem into simpler mixed integer linear programs by fixing $\mathbf{X}$ as constants and then focusing on two separate sets of decision variables, $\mathbf{Y}$ and $\mathbf{Z}$. 
For example, setting $\mathbf{X}_{jc}=\text{const}_{jc}$, the first quadratic constraint (Eq. \ref{eq:16}) simplifies into a linear constraint concerning $\mathbf{Y}_{ij}$:
\[
\mathbf{u}_j \geq A(\mathbf{Y}_{ij}(\text{const}_{ic}+\text{const}_{jc}-1)-1)+\mathbf{v}_i+\mathbf{e}_i.
\]
In a similar manner, the second and third quadratic constraints (Eqs. \ref{eq:20} and \ref{eq:21}) are transformed into linear constraints with respect to $\mathbf{Z}_{dl}$:
\[
\mathbf{v}_j \geq \max\{\mathbf{t}_d, \mathbf{u}_j\} + \sum_{l=1}^L\sum_{c=1}^Ctd^{(2)}(d,l,c)\text{const}_{jc}\mathbf{Z}_{dl}.
\]
We denote this induced MILP optimization problem as $\texttt{MILP} (\mathbf{Y},\mathbf{Z}|\mathbf{X})$. 
Similarly, when $\mathbf{Y}_{ij}$ and $\mathbf{Z}_{dl}$ are held as constants, the quadratic constraints become linear for the single decision variable $\mathbf{X}_{jc}$, leading to another MILP problem, which we denote as $\texttt{MILP}(\mathbf{X}|\mathbf{Y},\mathbf{Z})$. 
Essentially, these MILP problems with fixed decision variables can be viewed as conditional optimizations of the job assignment given the job precedence and data assignment, or vice versa.

\subsection{\name{}: Efficient Joint Optimization by Alternative MILP} \label{sec:3.2}

Building on the observations in Section \ref{sec:3.1}, \name{} efficiently addresses the quadratic constraints by alternately fixing subsets of decision variables and optimizing the others using MILP. 
Notably, this approach draws on the principles of a coordinate descent algorithm \citep{Wright2015CoordinateDA}, which has proven effective across various fields, including computed tomography \citep{sauer1993local}, protein structure prediction \citep{canutescu2003cyclic}, and large-scale optimization \citep{nesterov2012efficiency}.

In terms of algorithmic details, all decision variables are randomly initialized as $\mathbf{X}^{(0)}, \mathbf{Y}^{(0)}, \mathbf{Z}^{(0)}$ before optimization begins. 
During each iteration, \name{} first addresses $\texttt{MILP}(\mathbf{X}|\mathbf{Y},\mathbf{Z})$, finding the optimal solution $\mathbf{X}^{(t)}$ by fixing $\mathbf{Y}$ and $\mathbf{Z}$ with the values obtained from the previous iteration:
\begin{equation}
\label{eq:24}
\mathbf{X}^{(t)} \leftarrow \arg\min_{\mathbf{X}} (\texttt{MILP}(\mathbf{X}|\mathbf{Y}=\mathbf{Y}^{(t-1)}, \mathbf{Z}=\mathbf{Z}^{(t-1)})).
\end{equation}
Subsequently, \name{} tackles the other MILP problem, $\texttt{MILP}(\mathbf{Y},\mathbf{Z}|\mathbf{X})$, to determine $\mathbf{Y}^{(t)}$ and $\mathbf{Z}^{(t)}$ with $\mathbf{X}=\mathbf{X}^{(t)}$:
\begin{equation}
\label{eq:25}
\mathbf{Y}^{(t)},\mathbf{Z}^{(t)} \leftarrow \arg\min_{\mathbf{Y},\mathbf{Z}} (\texttt{MILP}(\mathbf{Y},\mathbf{Z}|\mathbf{X}=\mathbf{X}^{(t)})).
\end{equation}
It is important to note that the decision variables for each induced MILP problem are initialized with the solutions from the previous iteration to enhance optimization efficiency, i.e., $\mathbf{X} \leftarrow \mathbf{X}^{(t-1)}$ for Eq. \ref{eq:24} and $\mathbf{Y}, \mathbf{Z} \leftarrow \mathbf{Y}^{(t-1)}, \mathbf{Z}^{(t-1)}$ for Eq. \ref{eq:25}.

These alternating optimization steps are repeated for $T$ iterations to arrive at the final solution:
\[
\mathbf{X}^{*}, \mathbf{Y}^{*}, \mathbf{Z}^{*} \leftarrow \mathbf{X}^{(T)}, \mathbf{Y}^{(T)}, \mathbf{Z}^{(T)}.
\]
To resolve each induced MILP problem (Eqs. \ref{eq:24} and \ref{eq:25}), one can utilize existing solvers, such as SCIP \citep{achterberg2009scip}, CPLEX \citep{cplex2009v12}, or Gurobi \citep{gurobi}.

\section{Experiments}\label{sec:experiments}
\subsection{Experimental Setups}

\noindent\textbf{Simulated environment.} 
To evaluate the performance of {\name{}}, we simulate the CMS grid environment \citep{holtman2001cms}  following the prior works \citep{Phan05Co, TAHERI20131564, TAHERI20131885, casas2017balanced}. 
Specifically, the input data objects for each job are drawn from a Zipf distribution \citep{acdamic2012zipf}, and the simulated parameters of the environment are drawn independently from the uniform distributions, which are shown in Table \ref{tab:grid_setups}. To demonstrate the scalability of each algorithm, we consider three different simulated grids by varying the number of CNs, remote SNs, local SNs, data objects, and jobs.
We denote these setups as (1) Grid-small, (2) Grid-medium, (3) Grid-large, and the detailed information is also summarized in Table \ref{tab:grid_setups}. 
\begin{table}[h!]
\caption{Detailed characteristics of simulated grids.}\label{tab:grid_setups}
\begin{tabular*}{\tblwidth}{@{}L|C@{}C@{}C@{}}
\toprule
Attributes  & Grid-small & Grid-medium & Grid-large \\ 
\midrule
Number of CNs & 10 & 20 & 50\\
\midrule
Number of remote SNs & 10 & 20 & 50\\
\midrule
Number of local SNs & 10 & 20 & 50\\
\midrule
Number of jobs & 10 & 50 & 100\\
\midrule 
Number of data objects & 20 & 100 & 300\\
\midrule 
Object size & \multicolumn{3}{c}{50-1500 MB} \\
\midrule
WAN bandwidth & \multicolumn{3}{c}{700-1300    KB/s} \\ 
\midrule
LAN bandwidth &\multicolumn{3}{c}{7000-13000 KB/s}\\
\bottomrule
\end{tabular*}
\end{table}

\noindent\textbf{Baselines.} 
We compare 
\name{} with various baseline algorithms that can be grouped into three different categories. The first category includes a naive baseline, denoted by \textit{(a) Random}, which is based on a random initialization but without any optimization. 

The second group consists of \textit{independent} optimization algorithms, including
\textit{(b) MinTrans} \citep{tang2006impact}, which only optimizes data allocation through MILP with randomized job scheduling and job order;
\textit{(c) MinExe} \citep{ranganathan2002decoupling}, which only optimizes job assignment via MILP, based on randomized job order and data allocation;
\textit{(d) Greedy} \citep{Phan05Co}, which treats jobs 
in FIFO order and assigns each job to the next available CN; 
\textit{(e) Ensemble Greedy}, which randomizes the job order in the above greedy method to run it multiple times and picks the best run at the end.

The third group consists of several \textit{joint} optimization algorithms where all decision variables are optimized simultaneously: \textit{(f) GA} \citep{taheri2016genetic}, which progressively updates job scheduling and data allocation using a genetic algorithm \citep{bartz2014evolutionary};
\textit{(g) DIANA} \citep{mcclatchey2007data}, which categorizes jobs as computationally or data intensive.
For a computationally intensive job, DIANA migrates it to the CN with the lowest execution time; for a data intensive job, DIANA either migrates the job to the CN with the shortest data object downloading time;
\textit{(h) MIQP} \citep{Ko2019MIQP}, which directly solves the joint optimization problem via the existing MIQP solver and \textit{(i) JDS-HNN} \citep{TAHERI20131885}, which 
uses the Hopfield neural network to alternatively improve the current solution for job scheduling and data allocation, respectively.

\noindent\textbf{Implementation details.} 
For each method tested in a specific setup, we perform 10 independent experiments, each with a different set of environment parameters sampled from the distributions detailed above. The average makespan of these 10 experiments is presented in this section. We employ Gurobi \citep{gurobi} as the MIP solver. Additionally, for methods that depend on existing solvers (MinTrans, MinExe, MIQP, and \name{}), a consistent time budget is maintained throughout the optimization process. Specifically, we set time budgets as $B=3$, $30$, and $300$ (seconds) for the Grid-small, Grid-medium, and Grid-large configurations, respectively.

\subsection{Main Results}\label{sec:4.2}

\begin{table*}[h!]
\caption{Results on simulated grids: Grid-small, Grid-medium, Grid-large. The average makespan ($\downarrow$) over 10 independent experiments are reported (rows 2-4). The number below the makespan of each method, encapsulated by the parenthesis, indicates the relative improvement (\%) compared to Random baseline. Also, the average ranking  (Avg. Rank) across 3 test setups are reported in the last row. The best and second best results are highlighted with \textbf{bold} and \underline{underline}, respectively.}\label{tab:main_results}
\begin{tabular*}{\tblwidth}{@{}C|@{}C|@{}C@{}C@{}C@{}C|@{}C@{}C@{}C@{}C|@{}C}
\toprule
\multirow{2}{*}{Test setup} & \multirow{2}{*}{Random}  & \multicolumn{4}{c}{Independent Optimization} & \multicolumn{5}{c}{Joint Optimization} \\
 &  & MinTrans & MinExe & GA & Ens.Greedy  & GA & DIANA & MIQP  & JDS-HNN & AlterMILP  \\ 
\midrule
\multirow{2}{*}{Grid-small} & 2903 & 2819 & 2215 & 2278 & 1781  & 1875 & \underline{1736} & 2453 & 1914 & \textbf{1707} \\
& {\footnotesize (-0.00\%)} & {\footnotesize(-2.89\%)} & {\footnotesize(-23.7\%)} & {\footnotesize(-21.5\%)} & {\footnotesize(-38.6\%)}  & {\footnotesize(-35.4\%)} & {\footnotesize(-\underline{40.2}\%)} & {\footnotesize(-15.5\%)} & {\footnotesize(-34.1\%)} & {\footnotesize(\textbf{-41.2}\%)} \\ 
\multirow{2}{*}{Grid-medium} & 21052 & 19227 & \underline{9262} & 11304 & 10079  & 12122 & 63021 & \multirow{2}{*}{N/A} & 10221 & \textbf{8714} \\
& {\footnotesize (-0.00\%)} & {\footnotesize(-8.67\%)} & {\footnotesize(-\underline{56.0}\%)} & {\footnotesize(-46.3\%)} & {\footnotesize(-52.1\%)} & {\footnotesize(-42.4\%)} & {\footnotesize(+{199}\%)} &  & {\footnotesize(-51.4\%)} 
 & {\footnotesize(\textbf{-58.6}\%)} \\ 
\multirow{2}{*}{Grid-large} & 23221 & 18924 & \underline{8564} & 10371 & 9431  & 13222 & 121050 & \multirow{2}{*}{N/A} & 8951 & \textbf{7912} \\ 
& {\footnotesize (-0.00\%)} & {\footnotesize(-18.5\%)} & {\footnotesize(-\underline{63.1}\%)} & {\footnotesize(-55.3\%)} & {\footnotesize(-59.4\%)}  & {\footnotesize(-43.1\%)} & {\footnotesize(+{421}\%)} &  & {\footnotesize(-61.5\%)} & {\footnotesize(\textbf{-65.9}\%)} \\ 
\midrule
{Avg. Rank} & 8.7 & 7.7 & \underline{3.3} & 5.7 & \underline{3.3} & 5.3 & 6.7 & 9.3  & 4.0 & \textbf{1.0} \\
\bottomrule
\end{tabular*}
\end{table*}

Table \ref{tab:main_results} presents the experimental results across three simulated grids of varying sizes. For each method, we report the average makespan over 10 independent trials and the relative improvement compared to the Random baseline. Several key observations emerge from these results. 

Firstly, the effectiveness of the baseline methods significantly varies with the scale of the test setup. For instance, DIANA shows the second-best makespan in smaller setups (Grid-small), but its performance declines drastically in larger setups (Grid-medium and Grid-large), eventually performing worse than the Random baseline.

Secondly, MIQP exhibits extreme limitations in scalability. As MIQP seeks to solve the joint optimization problem directly through a conventional solver without modifications, it might initially seem like the most straightforward approach. However, as indicated in Table \ref{tab:main_results}, this method fails to find a feasible solution when the scale of the tested setup grows. We conjecture that this is because quadratic constraints incur a large solution space that exponentially
grows with respect to the number of optimization variables coming from three different sources (job assignment, job orders, and
data assignment).

The relative superiority of independent optimization methods over joint optimization methods highlights the inherent challenges posed by the expansive solution space, as demonstrated in Table \ref{tab:main_results}.  To offer a thorough comparison, we also include the average ranking (Avg. Rank) of various baseline algorithms across the three distinct setups in Table \ref{tab:main_results}. This analysis further underscores the effectiveness of simpler methods; for instance, independent optimization methods, such as MinExe and Ens. Greedy, achieve the second-lowest average rank.

In contrast, the proposed \name{} method effectively tackles this challenging problem, significantly outperforming other baseline approaches. Specifically, \name{} achieves a relative reduction of 55.2\% in makespan compared to the Random baseline, surpassing the second-best method, Ens. Greedy, by an additional 5.2\% improvement. Notably, \name{} consistently delivers the largest reduction in makespan across various test setups, demonstrating its reliable effectiveness in different scales of grid computing environments. It is important to note that the experiments with \name{} were conducted using fixed configurations for the time budget and the number of iterations ($T$), suggesting that further enhancements might be possible with additional hyper-parameter tuning.

\subsection{Additional Analyses with \name{}}\label{sec:4.3}

\begin{table}[h!]
\caption{Ablation study of AlterMILP on Grid-medium. \textit{JA}, \textit{JO}, \textit{DA} are abbreviations of jointly optimizing job assignment (\textit{JA}), job order (\textit{JO}), and data assignment (\textit{DA}), respectively. \textit{Iter} indicates the iterative updates with alternative optimization.}\label{tab:ours_ablation}
\begin{tabular*}{\tblwidth}{@{}L|@{}C@{}C@{}C@{}C|@{}C}
\toprule
\multirow{2}{*}{Methods} & \multicolumn{4}{c}{Configurations} & \multirow{2}{*}{Makespan} \\ 
  & \textit{JA} & \textit{JO} & \textit{DA} & \textit{Iter} &   \\ 
\midrule
MinExe & \textcolor{green}{\ding{51}} & \textcolor{red}{\ding{55}} & \textcolor{red}{\ding{55}} & \textcolor{red}{\ding{55}} & 9262 \\
MinTrans & \textcolor{red}{\ding{55}} & \textcolor{red}{\ding{55}} & \textcolor{green}{\ding{51}} & \textcolor{red}{\ding{55}} & 19227 \\
\midrule
\multirow{3}{*}{AlterMILP} & \textcolor{green}{\ding{51}} & \textcolor{red}{\ding{55}} & \textcolor{green}{\ding{51}} & \textcolor{red}{\ding{55}} & 9027  \\
& \textcolor{green}{\ding{51}} & \textcolor{green}{\ding{51}} & \textcolor{green}{\ding{51}} &\textcolor{red}{\ding{55}} & 8872  \\ 
 & \textcolor{green}{\ding{51}} & \textcolor{green}{\ding{51}} & \textcolor{green}{\ding{51}} & \textcolor{green}{\ding{51}} & \textbf{8714}  \\
\bottomrule
\end{tabular*}
\end{table}
\begin{figure*}[t]
\begin{center}
    {
    \subfigure[Grid-small]
        {
        \includegraphics[width=0.31\textwidth]{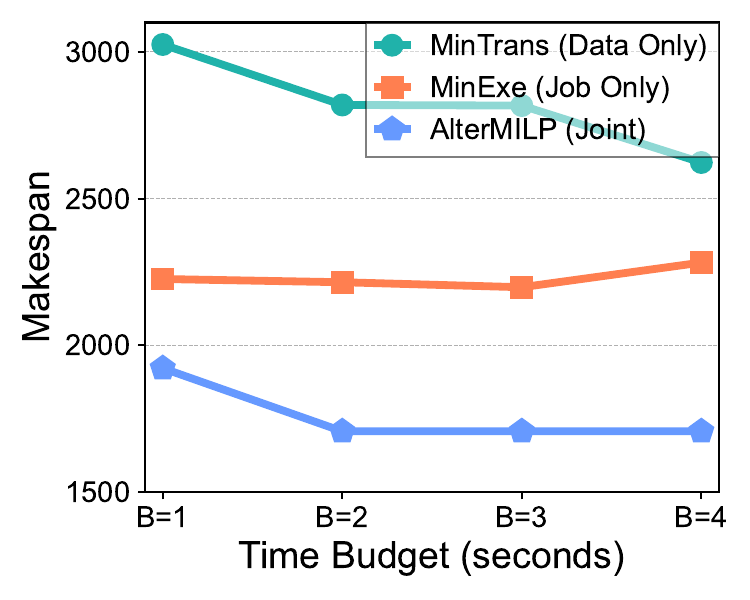}
        \label{fig:grid_small}
        }
    \subfigure[Grid-medium]
        {
        \includegraphics[width=0.31\textwidth]{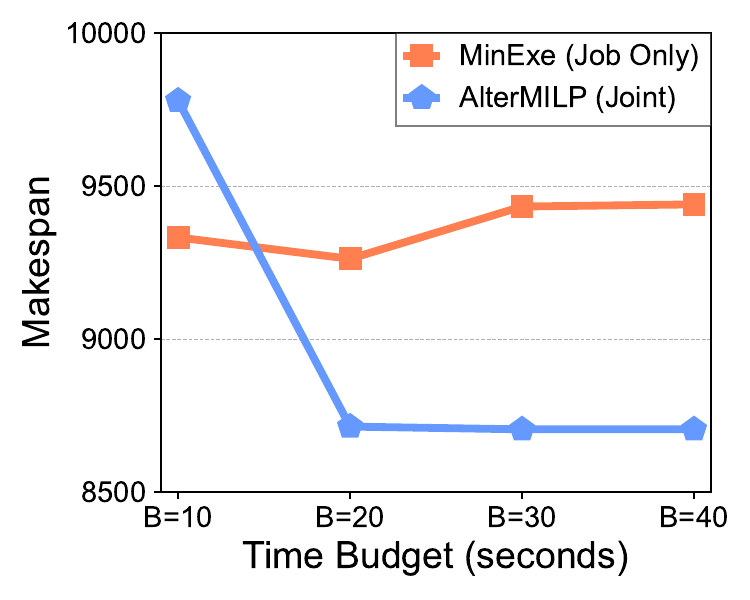}
        \label{fig:grid_medium}
        }
    \subfigure[Grid-large]
        {
        \includegraphics[width=0.31\textwidth]{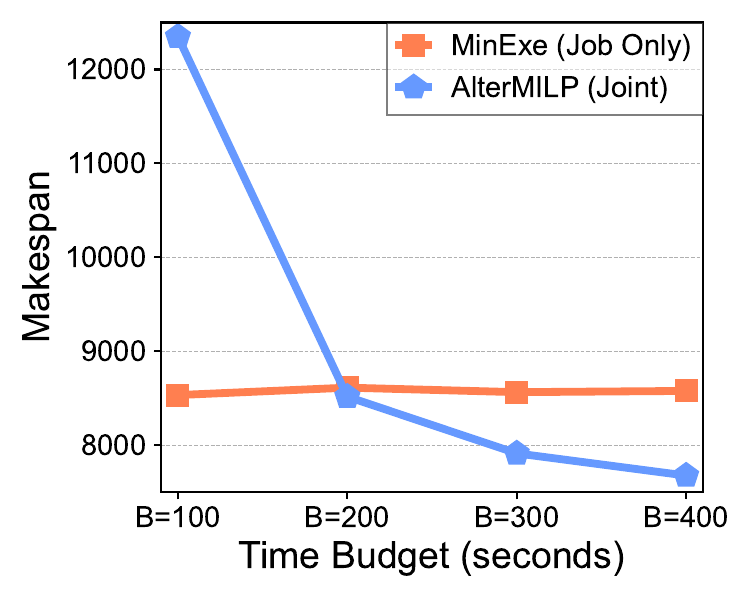}
        \label{fig:grid_large}
        } 
    }
\end{center}
\vspace{-0.2in}
\caption{Analyses of optimization-based methods (MinTrans, MinExe, AlterMILP) by varying time budget for the optimization. Since MinTrans largely underperforms other methods, we exclude it in Grid-medium/large for better visualization.}
\label{fig:analyses_time}
\end{figure*}
\begin{figure}[t!]
    \centering
    \includegraphics[width=0.9\linewidth]{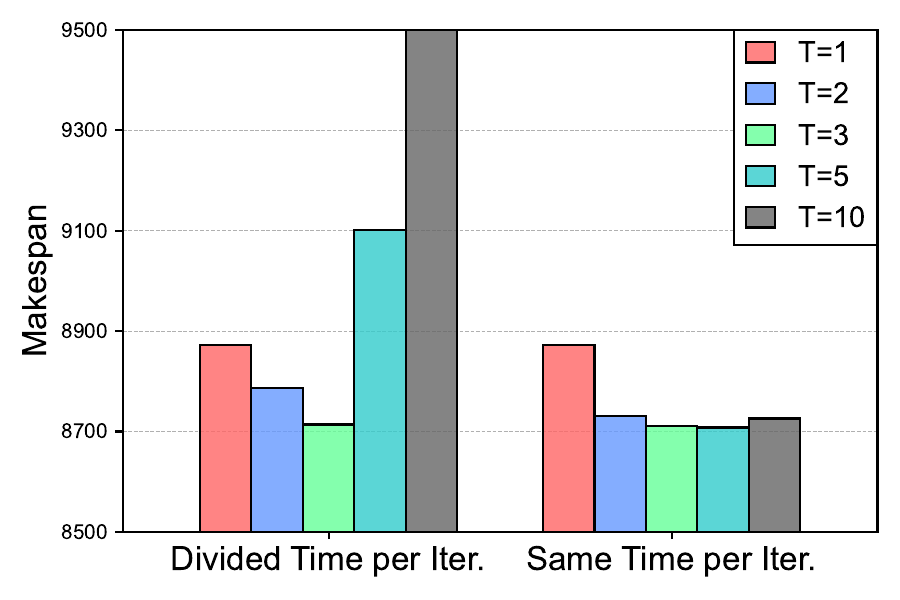}
    \vspace{-0.1in}
    \caption{Analyses of AlterMILP by varying the number of iterations $T$ on Grid-medium. (Left) \textit{Divided Time per Iter.}: total time budget $B$ is fixed, but $T$ is increased (i.e., less time per iteration). (Right) \textit{Same Time per Iter.}: using the same time per iteration, i.e., total time budget $B$ is enlarged with increased $T$.}\label{fig:analyses_iter}
\end{figure}

\noindent\textbf{Ablation Study.} To validate the effectiveness of the proposed components of \name{ in Section \ref{sec:method}, we perform ablation experiments on Grid-medium by accumulating the proposed components starting from the independent optimization methods, such as only optimizing job assignment (MinExe) or data assignment (MinTrans). 
The results are presented in Table \ref{tab:ours_ablation}.
Here, it can be observed that the use of alternative optimization is effective in solving the problem of joint job scheduling and data allocation. 
To be specific, when we alternatively update job assignment and data assignment, it yields 53.1\% (19227$ \rightarrow$ 9027) and 2.54\% (9262 $\rightarrow$ 9027) relative reduction in makespan compared to MinTrans and MinExe, respectively. 
In addition, when we incorporate the optimization of job order along with data assignment (Eq. \ref{eq:25}), the makespan is further reduced with a relative reduction of 1.72\% (9027 $\rightarrow$ 8872). 
Lastly, one can verify the effectiveness of iterating these alternative optimizations in the last row that yields 1.8\% of relative reduction in makespan (8872 $\rightarrow$ 8712); it confirms that the importance of adjusting the discrepancy under alternative optimizations occurs from the update of counterpart sub-problem.
Overall, the ablation study confirms that all the components proposed in \name{} clearly contribute to the reduction in makespan.

\noindent\textbf{Different time-budget to solve optimization.} 
We conduct additional experiments to further investigate the importance of the time budget to optimization-based methods like MinTrans, MinExe, and \name{}. Specifically, we adjust the time budget and set a time limit for our MILP solver. If the time limit is reached before the optimization is complete, the solver is terminated and the best solution found up to that point is returned as the final result. The outcomes of these experiments are displayed in Figure \ref{fig:analyses_time}.

Our results indicate that traditional optimization methods tend to converge more quickly than \name{}. In particular, MinExe consistently achieves a similar makespan even with a reduced time budget. In contrast, the performance of \name{} heavily depends on the available time budget, as the larger program it tries to optimize needs longer time for convergence. Consequently, \name{} would underperform MinExe with an insufficient time budget, as demonstrated in Figures \ref{fig:grid_medium} and \ref{fig:grid_large}. However, when enough time is provided, \name{} can achieve superior solutions that are not achievable by MinExe.

\noindent\textbf{Varying number of iterations with \name{}.} 
Finally, we examine the impact of the number of iterations $T$ on the performance of \name{}. To do this, we conduct a series of experiments with \name{} on the Grid-medium setup, varying $T$ across two distinct scenarios.

In the first scenario, we fix the total optimization time budget but increase $T$, which results in less time allocated per iteration. In the second scenario, we maintain the same time allocation per iteration, which allows for a larger total time budget as $T$ increases. The results of these experiments are shown in Figure \ref{fig:analyses_iter}.

From the results, it is clear that sufficient time per iteration is crucial. If this condition is not met (as seen in scenarios where $T=5$ and $10$, represented by the left bars in Figure \ref{fig:analyses_iter}), the performance of \name{} can suffer because the optimization process may be terminated early. Conversely, with adequate time per iteration, \name{} reliably converges with a smaller number of iterations ($T=3$). This suggests that using fewer iterations while allowing enough time for each can be an effective strategy to enhance \name{}'s performance.

\section{Related Work}
Grid computing environments have been extensively studied due to their potential to leverage distributed resources for high-performance computing tasks. A major challenge in these settings is optimizing job scheduling and data allocation, both crucial to improving resource utilization and reducing the makespan.

\subsection{Job Scheduling}
Job scheduling has been a central focus of grid computing research over the past few decades. Given its NP-hard nature, numerous heuristics have been developed to efficiently provide high-quality approximate solutions \citep{BRAUN2001810}. Notable examples include the MinExe \citep{ranganathan2002decoupling}, Min-min, and Max-min algorithms \citep{armstrong1998mapping, freund1998scheduling}.  These heuristics primarily focus on job execution time and machine availability, but often overlook the impact of data transfer time prior to job execution. To address this gap, context-aware methods have been proposed \citep{mcclatchey2007data, LI2016119, sahni2019data}, explicitly incorporating data transfer time into scheduling optimization. Despite this progress, these methods still assume static data availability, missing the optimization of the data allocation policy.

\subsection{Data Allocation}
Effective dynamic data allocation strategies have been shown to be significant in improving the efficiency of grid computing systems \citep{ALLCOCK2002749, bell2002simulation, tang2006impact, Venugopal2004grid}. Classical heuristics, such as Least Recently Used (LRU), Least Frequently Used (LFU) \citep{podlipnig2003cache}, and Latest Access Largest Weight (LALW) \citep{Ruay2008dynamic}, base data replication decisions on historical access patterns. In practice, data allocation strategies also need to balance storage overhead with access time, employing methods like adaptive replication \citep{fadaie2012replica} and hierarchical replication \citep{shorfuzzaman2010adaptive}. Additional strategies could also be developed to dynamically determine the number of data copies \citep{ranganathan2002improve}. In our study, we focus on a two-tier hierarchy of remote and local storage nodes, emphasizing full data replication with a single copy for each data object.

\subsection{Joint Optimization of Job Scheduling and Data Allocation}
Despite significant advances in optimizing job scheduling and data allocation separately, their joint optimization remains challenging due to their interdependence. Traditional approaches often address one aspect with simplifying assumptions and then fix that strategy when optimizing the other \citep{CHANG2007846,7874167}. This sequential approach can create a chicken-and-egg problem, leading to suboptimal performance of the initially optimized strategy. To overcome these limitations, meta-heuristics such as genetic algorithms \citep{Phan05Co}, artificial bee colony optimization \citep{TAHERI20131564}, and particle swarm optimization \citep{liu2013swarm} have been used for joint optimization. Recent studies have also begun to explore deep reinforcement learning for this purpose \citep{wei2021joint1, wei2021joint2, ZENG2024121}. Our work is closely related to Ko el al. \citep{Ko2019MIQP} and Govardhan et al. \citep{TAHERI20131885}. The former utilizes MIQP, 
which clearly suffers from scaling issues when the problems are large. The latter relies on a Hopfield neural network to optimize job scheduling and data allocation alternatively, similar to our approach in using the decomposition strategy. But either its way of decomposition or the optimization method is less efficient than ours in terms of optimization computation within a fixed budget.  In our experiments, we have found that \name{} significantly outperforms both methods, as shown in Table \ref{tab:main_results}.

\section{Conclusion}
In this work, we introduce AlterMILP, a method designed to jointly optimize job scheduling and data allocation in grid computing by alternatively optimizing a subset of variables via MILP. AlterMILP transforms the original nonlinear optimization problem into a series of mixed integer linear programs, solving each one effectively using an off-the-shelf MILP solver. Our results demonstrate that AlterMILP consistently outperforms other heuristics in simulated grid computing environments, achieving significant reductions in total makespan while exhibiting strong robustness.

Future research could enhance AlterMILP by incorporating additional decision variables, such as the number of replication copies and selective replication strategies. Furthermore, exploring diverse distributions for simulation parameters, beyond the current uniform distribution assumption, may provide deeper insights. We anticipate that future work will incorporate more realistic settings, enabling the application of AlterMILP to real-world systems.

\section*{Author contributions: CRediT}
\textbf{Shengyu Feng}: Conceptualization, Data curation, Investigation, Software, Writing – original draft. \textbf{Jaehyung Kim}: Formal analysis, Methodology, Software, Writing – original draft, Methodology. \textbf{Yiming Yang}: Funding acquisition, Supervision, Writing – review and editing. \textbf{Tadashi Maeno}: Project administration, Writing – review and editing. \textbf{Alexei Klimentov}: Funding acquisition, Project administration, Writing – review and editing. \textbf{Others}: Writing – review and editing.

\section*{Declaration of competing interest}
The authors declare that they have no known competing financial interests or personal relationships that could have appeared to influence the work reported in this paper.

\section*{Acknowledgement}
This material is based upon work supported by the U.S. Department of Energy,
Office of Science, Office of Advanced Scientific Computing Research under Award
Number DE-SC-0012704. This work was done in collaboration with the distributed
computing research and development program within the ATLAS Collaboration. We
thank our ATLAS colleagues, and in particular, the contributions of the ATLAS
Distributed Computing team for their support. We would also like to express our
deepest gratitude to Prof.~Kaushik De at University of Texas at Arlington.

\section*{Data availability}
All data used in this work is simulated and the details are included in Section \ref{sec:experiments}.

\section*{Declaration of generative AI and AI-assisted technologies in the writing process}

During the preparation of this work the author(s) used GPT in order to polish the writing. After using this tool/service, the author(s) reviewed and edited the content as needed and take(s) full responsibility for the content of the publication.

\bibliographystyle{cas-model2-names}

\bibliography{main}

\begin{thebibliography}{50}
\expandafter\ifx\csname natexlab\endcsname\relax\def\natexlab#1{#1}\fi
\providecommand{\url}[1]{\texttt{#1}}
\providecommand{\href}[2]{#2}
\providecommand{\path}[1]{#1}
\providecommand{\DOIprefix}{doi:}
\providecommand{\ArXivprefix}{arXiv:}
\providecommand{\URLprefix}{URL: }
\providecommand{\Pubmedprefix}{pmid:}
\providecommand{\doi}[1]{\href{http://dx.doi.org/#1}{\path{#1}}}
\providecommand{\Pubmed}[1]{\href{pmid:#1}{\path{#1}}}
\providecommand{\bibinfo}[2]{#2}
\ifx\xfnm\relax \def\xfnm[#1]{\unskip,\space#1}\fi
\bibitem[{Achterberg(2009)}]{achterberg2009scip}
\bibinfo{author}{Achterberg, T.}, \bibinfo{year}{2009}.
\newblock \bibinfo{title}{Scip: solving constraint integer programs}.
\newblock \bibinfo{journal}{Mathematical Programming Computation} \bibinfo{volume}{1}, \bibinfo{pages}{1--41}.
\bibitem[{Adamic(2012)}]{acdamic2012zipf}
\bibinfo{author}{Adamic, L.}, \bibinfo{year}{2012}.
\newblock \bibinfo{title}{Zipf, power-laws, and pareto- a ranking tutorial} .
\bibitem[{Allcock et~al.(2002)Allcock, Bester, Bresnahan, Chervenak, Foster, Kesselman, Meder, Nefedova, Quesnel and Tuecke}]{ALLCOCK2002749}
\bibinfo{author}{Allcock, B.}, \bibinfo{author}{Bester, J.}, \bibinfo{author}{Bresnahan, J.}, \bibinfo{author}{Chervenak, A.L.}, \bibinfo{author}{Foster, I.}, \bibinfo{author}{Kesselman, C.}, \bibinfo{author}{Meder, S.}, \bibinfo{author}{Nefedova, V.}, \bibinfo{author}{Quesnel, D.}, \bibinfo{author}{Tuecke, S.}, \bibinfo{year}{2002}.
\newblock \bibinfo{title}{Data management and transfer in high-performance computational grid environments}.
\newblock \bibinfo{journal}{Parallel Computing} \bibinfo{volume}{28}, \bibinfo{pages}{749--771}.
\newblock \URLprefix \url{https://www.sciencedirect.com/science/article/pii/S0167819102000947}, \DOIprefix\doi{https://doi.org/10.1016/S0167-8191(02)00094-7}.
\bibitem[{Amjad et~al.(2012)Amjad, Sher and Daud}]{AMJAD2012337}
\bibinfo{author}{Amjad, T.}, \bibinfo{author}{Sher, M.}, \bibinfo{author}{Daud, A.}, \bibinfo{year}{2012}.
\newblock \bibinfo{title}{A survey of dynamic replication strategies for improving data availability in data grids}.
\newblock \bibinfo{journal}{Future Generation Computer Systems} \bibinfo{volume}{28}, \bibinfo{pages}{337--349}.
\newblock \URLprefix \url{https://www.sciencedirect.com/science/article/pii/S0167739X11001208}, \DOIprefix\doi{https://doi.org/10.1016/j.future.2011.06.009}.
\bibitem[{Armstrong et~al.(1998)Armstrong, Hensgen and Kidd}]{armstrong1998mapping}
\bibinfo{author}{Armstrong, R.}, \bibinfo{author}{Hensgen, D.}, \bibinfo{author}{Kidd, T.}, \bibinfo{year}{1998}.
\newblock \bibinfo{title}{The relative performance of various mapping algorithms is independent of sizable variances in run-time predictions}, in: \bibinfo{booktitle}{Proceedings Seventh Heterogeneous Computing Workshop (HCW'98)}, pp. \bibinfo{pages}{79--87}.
\newblock \DOIprefix\doi{10.1109/HCW.1998.666547}.
\bibitem[{Au(2009)}]{Au2009Grid}
\bibinfo{author}{Au, L.}, \bibinfo{year}{2009}.
\newblock \bibinfo{title}{Grid computing for bioinformatics and computational biology}.
\newblock \bibinfo{journal}{Quarterly Review of Biology - QUART REV BIOL} \bibinfo{volume}{84}, \bibinfo{pages}{85--86}.
\newblock \DOIprefix\doi{10.1086/598260}.
\bibitem[{Bartz-Beielstein et~al.(2014)Bartz-Beielstein, Branke, Mehnen and Mersmann}]{bartz2014evolutionary}
\bibinfo{author}{Bartz-Beielstein, T.}, \bibinfo{author}{Branke, J.}, \bibinfo{author}{Mehnen, J.}, \bibinfo{author}{Mersmann, O.}, \bibinfo{year}{2014}.
\newblock \bibinfo{title}{Evolutionary algorithms}.
\newblock \bibinfo{journal}{Wiley Interdisciplinary Reviews: Data Mining and Knowledge Discovery} \bibinfo{volume}{4}, \bibinfo{pages}{178--195}.
\bibitem[{Bell et~al.(2002)Bell, Cameron, Capozza, Millar, Stockinger and Zini}]{bell2002simulation}
\bibinfo{author}{Bell, W.H.}, \bibinfo{author}{Cameron, D.G.}, \bibinfo{author}{Capozza, L.}, \bibinfo{author}{Millar, A.P.}, \bibinfo{author}{Stockinger, K.}, \bibinfo{author}{Zini, F.}, \bibinfo{year}{2002}.
\newblock \bibinfo{title}{Simulation of dynamic grid replication strategies in optorsim}, in: \bibinfo{editor}{Parashar, M.} (Ed.), \bibinfo{booktitle}{Grid Computing --- GRID 2002}, \bibinfo{publisher}{Springer Berlin Heidelberg}, \bibinfo{address}{Berlin, Heidelberg}. pp. \bibinfo{pages}{46--57}.
\bibitem[{Benacchio and Pasian(2007)}]{Benacchio07Grid}
\bibinfo{author}{Benacchio, L.}, \bibinfo{author}{Pasian, F.}, \bibinfo{year}{2007}.
\newblock \bibinfo{title}{Grid-enabled astrophysics} .
\bibitem[{Braun et~al.(2001)Braun, Siegel, Beck, Bölöni, Maheswaran, Reuther, Robertson, Theys, Yao, Hensgen and Freund}]{BRAUN2001810}
\bibinfo{author}{Braun, T.D.}, \bibinfo{author}{Siegel, H.J.}, \bibinfo{author}{Beck, N.}, \bibinfo{author}{Bölöni, L.L.}, \bibinfo{author}{Maheswaran, M.}, \bibinfo{author}{Reuther, A.I.}, \bibinfo{author}{Robertson, J.P.}, \bibinfo{author}{Theys, M.D.}, \bibinfo{author}{Yao, B.}, \bibinfo{author}{Hensgen, D.}, \bibinfo{author}{Freund, R.F.}, \bibinfo{year}{2001}.
\newblock \bibinfo{title}{A comparison of eleven static heuristics for mapping a class of independent tasks onto heterogeneous distributed computing systems}.
\newblock \bibinfo{journal}{Journal of Parallel and Distributed Computing} \bibinfo{volume}{61}, \bibinfo{pages}{810--837}.
\newblock \URLprefix \url{https://www.sciencedirect.com/science/article/pii/S0743731500917143}, \DOIprefix\doi{https://doi.org/10.1006/jpdc.2000.1714}.
\bibitem[{Canutescu and Dunbrack~Jr(2003)}]{canutescu2003cyclic}
\bibinfo{author}{Canutescu, A.A.}, \bibinfo{author}{Dunbrack~Jr, R.L.}, \bibinfo{year}{2003}.
\newblock \bibinfo{title}{Cyclic coordinate descent: A robotics algorithm for protein loop closure}.
\newblock \bibinfo{journal}{Protein science} \bibinfo{volume}{12}, \bibinfo{pages}{963--972}.
\bibitem[{Casas et~al.(2017)Casas, Taheri, Ranjan, Wang and Zomaya}]{casas2017balanced}
\bibinfo{author}{Casas, I.}, \bibinfo{author}{Taheri, J.}, \bibinfo{author}{Ranjan, R.}, \bibinfo{author}{Wang, L.}, \bibinfo{author}{Zomaya, A.Y.}, \bibinfo{year}{2017}.
\newblock \bibinfo{title}{A balanced scheduler with data reuse and replication for scientific workflows in cloud computing systems}.
\newblock \bibinfo{journal}{Future Generation Computer Systems} \bibinfo{volume}{74}, \bibinfo{pages}{168--178}.
\bibitem[{Chang et~al.(2008)Chang, Chang and Wang}]{Ruay2008dynamic}
\bibinfo{author}{Chang, R.S.}, \bibinfo{author}{Chang, H.P.}, \bibinfo{author}{Wang, Y.T.}, \bibinfo{year}{2008}.
\newblock \bibinfo{title}{A dynamic weighted data replication strategy in data grids}, in: \bibinfo{booktitle}{2008 IEEE/ACS International Conference on Computer Systems and Applications}, pp. \bibinfo{pages}{414--421}.
\newblock \DOIprefix\doi{10.1109/AICCSA.2008.4493567}.
\bibitem[{Chang et~al.(2007)Chang, Chang and Lin}]{CHANG2007846}
\bibinfo{author}{Chang, R.S.}, \bibinfo{author}{Chang, J.S.}, \bibinfo{author}{Lin, S.Y.}, \bibinfo{year}{2007}.
\newblock \bibinfo{title}{Job scheduling and data replication on data grids}.
\newblock \bibinfo{journal}{Future Generation Computer Systems} \bibinfo{volume}{23}, \bibinfo{pages}{846--860}.
\newblock \URLprefix \url{https://www.sciencedirect.com/science/article/pii/S0167739X07000301}, \DOIprefix\doi{https://doi.org/10.1016/j.future.2007.02.008}.
\bibitem[{Cplex(2009)}]{cplex2009v12}
\bibinfo{author}{Cplex, I.I.}, \bibinfo{year}{2009}.
\newblock \bibinfo{title}{V12. 1: User’s manual for cplex}.
\newblock \bibinfo{journal}{International Business Machines Corporation} \bibinfo{volume}{46}, \bibinfo{pages}{157}.
\bibitem[{Du and Leung(1989)}]{Du1989Complexity}
\bibinfo{author}{Du, J.}, \bibinfo{author}{Leung, J.Y.T.}, \bibinfo{year}{1989}.
\newblock \bibinfo{title}{Complexity of scheduling parallel task systems}.
\newblock \bibinfo{journal}{SIAM Journal on Discrete Mathematics} \bibinfo{volume}{2}, \bibinfo{pages}{473--487}.
\newblock \URLprefix \url{https://doi.org/10.1137/0402042}, \DOIprefix\doi{10.1137/0402042}, \href{http://arxiv.org/abs/https://doi.org/10.1137/0402042}{\tt arXiv:https://doi.org/10.1137/0402042}.
\bibitem[{Evans(2009)}]{Evans2009TheLH}
\bibinfo{author}{Evans, L.R.}, \bibinfo{year}{2009}.
\newblock \bibinfo{title}{The large hadron collider : a marvel of technology}.
\newblock \URLprefix \url{https://api.semanticscholar.org/CorpusID:118322688}.
\bibitem[{Fadaie and Rahmani(2012)}]{fadaie2012replica}
\bibinfo{author}{Fadaie, Z.}, \bibinfo{author}{Rahmani, A.}, \bibinfo{year}{2012}.
\newblock \bibinfo{title}{A new replica placement algorithm in data grid}.
\newblock \bibinfo{journal}{Int J Comput Sci Issues} \bibinfo{volume}{9}, \bibinfo{pages}{491--507}.
\bibitem[{Freund et~al.(1998)Freund, Gherrity, Ambrosius, Campbell, Halderman, Hensgen, Keith, Kidd, Kussow, Lima, Mirabile, Moore, Rust and Siegel}]{freund1998scheduling}
\bibinfo{author}{Freund, R.}, \bibinfo{author}{Gherrity, M.}, \bibinfo{author}{Ambrosius, S.}, \bibinfo{author}{Campbell, M.}, \bibinfo{author}{Halderman, M.}, \bibinfo{author}{Hensgen, D.}, \bibinfo{author}{Keith, E.}, \bibinfo{author}{Kidd, T.}, \bibinfo{author}{Kussow, M.}, \bibinfo{author}{Lima, J.}, \bibinfo{author}{Mirabile, F.}, \bibinfo{author}{Moore, L.}, \bibinfo{author}{Rust, B.}, \bibinfo{author}{Siegel, H.}, \bibinfo{year}{1998}.
\newblock \bibinfo{title}{Scheduling resources in multi-user, heterogeneous, computing environments with smartnet}, in: \bibinfo{booktitle}{Proceedings Seventh Heterogeneous Computing Workshop (HCW'98)}, pp. \bibinfo{pages}{184--199}.
\newblock \DOIprefix\doi{10.1109/HCW.1998.666558}.
\bibitem[{Govardhan and Dugyani(2024)}]{Govardhan2024survey}
\bibinfo{author}{Govardhan, D.}, \bibinfo{author}{Dugyani, R.}, \bibinfo{year}{2024}.
\newblock \bibinfo{title}{Survey on data replication in cloud systems}.
\newblock \bibinfo{journal}{Web Intelligence} \bibinfo{volume}{22}, \bibinfo{pages}{1--27}.
\newblock \DOIprefix\doi{10.3233/WEB-230087}.
\bibitem[{{Gurobi Optimization, LLC}(2023)}]{gurobi}
\bibinfo{author}{{Gurobi Optimization, LLC}}, \bibinfo{year}{2023}.
\newblock \bibinfo{title}{{Gurobi Optimizer Reference Manual}}.
\newblock \URLprefix \url{https://www.gurobi.com}.
\bibitem[{Holtman(2001)}]{holtman2001cms}
\bibinfo{author}{Holtman, K.}, \bibinfo{year}{2001}.
\newblock \bibinfo{title}{Cms data grid system - overview and requirements} .
\bibitem[{Hussain et~al.(2013)Hussain, Malik, Hameed, Khan, Bickler, Min-Allah, Qureshi, Zhang, Yongji, Ghani, Kolodziej, Zomaya, Xu, Balaji, Vishnu, Pinel, Pecero, Kliazovich, Bouvry, Li, Wang, Chen and Rayes}]{HUSSAIN2013709}
\bibinfo{author}{Hussain, H.}, \bibinfo{author}{Malik, S.U.R.}, \bibinfo{author}{Hameed, A.}, \bibinfo{author}{Khan, S.U.}, \bibinfo{author}{Bickler, G.}, \bibinfo{author}{Min-Allah, N.}, \bibinfo{author}{Qureshi, M.B.}, \bibinfo{author}{Zhang, L.}, \bibinfo{author}{Yongji, W.}, \bibinfo{author}{Ghani, N.}, \bibinfo{author}{Kolodziej, J.}, \bibinfo{author}{Zomaya, A.Y.}, \bibinfo{author}{Xu, C.Z.}, \bibinfo{author}{Balaji, P.}, \bibinfo{author}{Vishnu, A.}, \bibinfo{author}{Pinel, F.}, \bibinfo{author}{Pecero, J.E.}, \bibinfo{author}{Kliazovich, D.}, \bibinfo{author}{Bouvry, P.}, \bibinfo{author}{Li, H.}, \bibinfo{author}{Wang, L.}, \bibinfo{author}{Chen, D.}, \bibinfo{author}{Rayes, A.}, \bibinfo{year}{2013}.
\newblock \bibinfo{title}{A survey on resource allocation in high performance distributed computing systems}.
\newblock \bibinfo{journal}{Parallel Computing} \bibinfo{volume}{39}, \bibinfo{pages}{709--736}.
\newblock \URLprefix \url{https://www.sciencedirect.com/science/article/pii/S016781911300121X}, \DOIprefix\doi{https://doi.org/10.1016/j.parco.2013.09.009}.
\bibitem[{Ko et~al.(2019)Ko, Kang and Joo}]{Ko2019MIQP}
\bibinfo{author}{Ko, R.}, \bibinfo{author}{Kang, D.}, \bibinfo{author}{Joo, S.K.}, \bibinfo{year}{2019}.
\newblock \bibinfo{title}{Mixed integer quadratic programming based scheduling methods for day-ahead bidding and intra-day operation of virtual power plant}.
\newblock \bibinfo{journal}{Energies} \bibinfo{volume}{12}, \bibinfo{pages}{1410}.
\newblock \DOIprefix\doi{10.3390/en12081410}.
\bibitem[{Li et~al.(2016)Li, Jiang and Ruiz}]{LI2016119}
\bibinfo{author}{Li, X.}, \bibinfo{author}{Jiang, T.}, \bibinfo{author}{Ruiz, R.}, \bibinfo{year}{2016}.
\newblock \bibinfo{title}{Heuristics for periodical batch job scheduling in a mapreduce computing framework}.
\newblock \bibinfo{journal}{Information Sciences} \bibinfo{volume}{326}, \bibinfo{pages}{119--133}.
\newblock \URLprefix \url{https://www.sciencedirect.com/science/article/pii/S0020025515005459}, \DOIprefix\doi{https://doi.org/10.1016/j.ins.2015.07.040}.
\bibitem[{Liu et~al.(2013)Liu, Abraham, Snasel and Mcloone}]{liu2013swarm}
\bibinfo{author}{Liu, H.}, \bibinfo{author}{Abraham, A.}, \bibinfo{author}{Snasel, V.}, \bibinfo{author}{Mcloone, S.}, \bibinfo{year}{2013}.
\newblock \bibinfo{title}{Swarm scheduling approaches for work-flow applications with security constraints in distributed data-intensive computing environments}.
\newblock \bibinfo{journal}{Information Sciences} \bibinfo{volume}{192}, \bibinfo{pages}{228--243}.
\newblock \DOIprefix\doi{10.1016/j.ins.2011.12.032}.
\bibitem[{McClatchey et~al.(2007)McClatchey, Anjum, Stockinger, Ali, Willers and Thomas}]{mcclatchey2007data}
\bibinfo{author}{McClatchey, R.}, \bibinfo{author}{Anjum, A.}, \bibinfo{author}{Stockinger, H.}, \bibinfo{author}{Ali, A.}, \bibinfo{author}{Willers, I.}, \bibinfo{author}{Thomas, M.}, \bibinfo{year}{2007}.
\newblock \bibinfo{title}{Data intensive and network aware (diana) grid scheduling}.
\newblock \bibinfo{journal}{Journal of Grid computing} \bibinfo{volume}{5}, \bibinfo{pages}{43--64}.
\bibitem[{Mishra et~al.(2014)Mishra, Patel, Rout and Mund}]{Mishra2014Survey}
\bibinfo{author}{Mishra, M.}, \bibinfo{author}{Patel, Y.}, \bibinfo{author}{Rout, Y.}, \bibinfo{author}{Mund, G.}, \bibinfo{year}{2014}.
\newblock \bibinfo{title}{A survey on scheduling heuristics in grid computing environment}.
\newblock \bibinfo{journal}{International Journal of Modern Education and Computer Science} \bibinfo{volume}{6}, \bibinfo{pages}{57--83}.
\newblock \DOIprefix\doi{10.5815/ijmecs.2014.10.08}.
\bibitem[{Nesterov(2012)}]{nesterov2012efficiency}
\bibinfo{author}{Nesterov, Y.}, \bibinfo{year}{2012}.
\newblock \bibinfo{title}{Efficiency of coordinate descent methods on huge-scale optimization problems}.
\newblock \bibinfo{journal}{SIAM Journal on Optimization} \bibinfo{volume}{22}, \bibinfo{pages}{341--362}.
\bibitem[{Phan et~al.(2005)Phan, Ranganathan and Sion}]{Phan05Co}
\bibinfo{author}{Phan, T.}, \bibinfo{author}{Ranganathan, K.}, \bibinfo{author}{Sion, R.}, \bibinfo{year}{2005}.
\newblock \bibinfo{title}{Evolving toward the perfect schedule: Co-scheduling job assignments and data replication in wide-area systems using a genetic algorithm}, in: \bibinfo{editor}{Feitelson, D.}, \bibinfo{editor}{Frachtenberg, E.}, \bibinfo{editor}{Rudolph, L.}, \bibinfo{editor}{Schwiegelshohn, U.} (Eds.), \bibinfo{booktitle}{Job Scheduling Strategies for Parallel Processing}, \bibinfo{publisher}{Springer Berlin Heidelberg}, \bibinfo{address}{Berlin, Heidelberg}. pp. \bibinfo{pages}{173--193}.
\bibitem[{Podlipnig and B\"{o}sz\"{o}rmenyi(2003)}]{podlipnig2003cache}
\bibinfo{author}{Podlipnig, S.}, \bibinfo{author}{B\"{o}sz\"{o}rmenyi, L.}, \bibinfo{year}{2003}.
\newblock \bibinfo{title}{A survey of web cache replacement strategies}.
\newblock \bibinfo{journal}{ACM Comput. Surv.} \bibinfo{volume}{35}, \bibinfo{pages}{374–398}.
\newblock \URLprefix \url{https://doi.org/10.1145/954339.954341}, \DOIprefix\doi{10.1145/954339.954341}.
\bibitem[{Ranganathan and Foster(2002)}]{ranganathan2002decoupling}
\bibinfo{author}{Ranganathan, K.}, \bibinfo{author}{Foster, I.}, \bibinfo{year}{2002}.
\newblock \bibinfo{title}{Decoupling computation and data scheduling in distributed data-intensive applications}, in: \bibinfo{booktitle}{Proceedings 11th IEEE International Symposium on High Performance Distributed Computing}, \bibinfo{organization}{IEEE}. pp. \bibinfo{pages}{352--358}.
\bibitem[{Ranganathan et~al.(2002)Ranganathan, Iamnitchi and Foster}]{ranganathan2002improve}
\bibinfo{author}{Ranganathan, K.}, \bibinfo{author}{Iamnitchi, A.}, \bibinfo{author}{Foster, I.}, \bibinfo{year}{2002}.
\newblock \bibinfo{title}{Improving data availability through dynamic model-driven replication in large peer-to-peer communities}, in: \bibinfo{booktitle}{2nd IEEE/ACM International Symposium on Cluster Computing and the Grid (CCGRID'02)}, pp. \bibinfo{pages}{376--376}.
\newblock \DOIprefix\doi{10.1109/CCGRID.2002.1017164}.
\bibitem[{Renard et~al.(2009)Renard, Badoux, Petitdidier and Cossu}]{Renard09Grid}
\bibinfo{author}{Renard, P.}, \bibinfo{author}{Badoux, V.}, \bibinfo{author}{Petitdidier, M.}, \bibinfo{author}{Cossu, R.}, \bibinfo{year}{2009}.
\newblock \bibinfo{title}{Grid computing for earth science}.
\newblock \bibinfo{journal}{Eos, Transactions American Geophysical Union} \bibinfo{volume}{90}.
\newblock \DOIprefix\doi{10.1029/2009EO140002}.
\bibitem[{Sahni et~al.(2019)Sahni, Cao and Yang}]{sahni2019data}
\bibinfo{author}{Sahni, Y.}, \bibinfo{author}{Cao, J.}, \bibinfo{author}{Yang, L.}, \bibinfo{year}{2019}.
\newblock \bibinfo{title}{Data-aware task allocation for achieving low latency in collaborative edge computing}.
\newblock \bibinfo{journal}{IEEE Internet of Things Journal} \bibinfo{volume}{6}, \bibinfo{pages}{3512--3524}.
\newblock \DOIprefix\doi{10.1109/JIOT.2018.2886757}.
\bibitem[{Sauer and Bouman(1993)}]{sauer1993local}
\bibinfo{author}{Sauer, K.}, \bibinfo{author}{Bouman, C.}, \bibinfo{year}{1993}.
\newblock \bibinfo{title}{A local update strategy for iterative reconstruction from projections}.
\newblock \bibinfo{journal}{IEEE Transactions on Signal Processing} \bibinfo{volume}{41}, \bibinfo{pages}{534--548}.
\bibitem[{Shmoys and Tardos(1993)}]{Shmoys1993Approximation}
\bibinfo{author}{Shmoys, D.}, \bibinfo{author}{Tardos, E.}, \bibinfo{year}{1993}.
\newblock \bibinfo{title}{Approximation algorithm for the generalized assignment problem}.
\newblock \bibinfo{journal}{Math. Program.} \bibinfo{volume}{62}, \bibinfo{pages}{461--474}.
\newblock \DOIprefix\doi{10.1007/BF01585178}.
\bibitem[{Shorfuzzaman et~al.(2010)Shorfuzzaman, Graham and Eskicioglu}]{shorfuzzaman2010adaptive}
\bibinfo{author}{Shorfuzzaman, M.}, \bibinfo{author}{Graham, P.}, \bibinfo{author}{Eskicioglu, R.}, \bibinfo{year}{2010}.
\newblock \bibinfo{title}{Adaptive replica placement in hierarchical data grids}.
\newblock \bibinfo{journal}{Journal of Physics: Conference Series} \bibinfo{volume}{256}, \bibinfo{pages}{012020}.
\newblock \DOIprefix\doi{10.1088/1742-6596/256/1/012020}.
\bibitem[{Taheri et~al.(2013a)Taheri, {Choon Lee}, Zomaya and Siegel}]{TAHERI20131564}
\bibinfo{author}{Taheri, J.}, \bibinfo{author}{{Choon Lee}, Y.}, \bibinfo{author}{Zomaya, A.Y.}, \bibinfo{author}{Siegel, H.J.}, \bibinfo{year}{2013}a.
\newblock \bibinfo{title}{A bee colony based optimization approach for simultaneous job scheduling and data replication in grid environments}.
\newblock \bibinfo{journal}{Computers \& Operations Research} \bibinfo{volume}{40}, \bibinfo{pages}{1564--1578}.
\newblock \URLprefix \url{https://www.sciencedirect.com/science/article/pii/S0305054811003376}, \DOIprefix\doi{https://doi.org/10.1016/j.cor.2011.11.012}. \bibinfo{note}{emergent Nature Inspired Algorithms for Multi-Objective Optimization}.
\bibitem[{Taheri et~al.(2013b)Taheri, Zomaya, Bouvry and Khan}]{TAHERI20131885}
\bibinfo{author}{Taheri, J.}, \bibinfo{author}{Zomaya, A.Y.}, \bibinfo{author}{Bouvry, P.}, \bibinfo{author}{Khan, S.U.}, \bibinfo{year}{2013}b.
\newblock \bibinfo{title}{Hopfield neural network for simultaneous job scheduling and data replication in grids}.
\newblock \bibinfo{journal}{Future Generation Computer Systems} \bibinfo{volume}{29}, \bibinfo{pages}{1885--1900}.
\newblock \URLprefix \url{https://www.sciencedirect.com/science/article/pii/S0167739X13000800}, \DOIprefix\doi{https://doi.org/10.1016/j.future.2013.04.020}. \bibinfo{note}{including Special sections: Advanced Cloud Monitoring Systems \& The fourth IEEE International Conference on e-Science 2011 — e-Science Applications and Tools \& Cluster, Grid, and Cloud Computing}.
\bibitem[{Taheri et~al.(2016)Taheri, Zomaya and Khan}]{taheri2016genetic}
\bibinfo{author}{Taheri, J.}, \bibinfo{author}{Zomaya, A.Y.}, \bibinfo{author}{Khan, S.U.}, \bibinfo{year}{2016}.
\newblock \bibinfo{title}{Genetic algorithm in finding pareto frontier of optimizing data transfer versus job execution in grids}.
\newblock \bibinfo{journal}{Concurrency and Computation: Practice and Experience} \bibinfo{volume}{28}, \bibinfo{pages}{1715--1736}.
\bibitem[{Tang et~al.(2017)Tang, Chen, Hefferman, Pei, Wei, He and Yang}]{7874167}
\bibinfo{author}{Tang, B.}, \bibinfo{author}{Chen, Z.}, \bibinfo{author}{Hefferman, G.}, \bibinfo{author}{Pei, S.}, \bibinfo{author}{Wei, T.}, \bibinfo{author}{He, H.}, \bibinfo{author}{Yang, Q.}, \bibinfo{year}{2017}.
\newblock \bibinfo{title}{Incorporating intelligence in fog computing for big data analysis in smart cities}.
\newblock \bibinfo{journal}{IEEE Transactions on Industrial Informatics} \bibinfo{volume}{13}, \bibinfo{pages}{2140--2150}.
\newblock \DOIprefix\doi{10.1109/TII.2017.2679740}.
\bibitem[{Tang et~al.(2006)Tang, Lee, Tang and Yeo}]{tang2006impact}
\bibinfo{author}{Tang, M.}, \bibinfo{author}{Lee, B.S.}, \bibinfo{author}{Tang, X.}, \bibinfo{author}{Yeo, C.K.}, \bibinfo{year}{2006}.
\newblock \bibinfo{title}{The impact of data replication on job scheduling performance in the data grid}.
\newblock \bibinfo{journal}{Future Generation Computer Systems} \bibinfo{volume}{22}, \bibinfo{pages}{254--268}.
\bibitem[{Tyagi and Gupta(2018)}]{Tyagi2018Survey}
\bibinfo{author}{Tyagi, R.}, \bibinfo{author}{Gupta, S.}, \bibinfo{year}{2018}.
\newblock \bibinfo{title}{A Survey on Scheduling Algorithms for Parallel and Distributed Systems}.
\newblock pp. \bibinfo{pages}{51--64}.
\newblock \DOIprefix\doi{10.1007/978-981-10-7656-5_7}.
\bibitem[{Venugopal et~al.(2004)Venugopal, Buyya and Winton}]{Venugopal2004grid}
\bibinfo{author}{Venugopal, S.}, \bibinfo{author}{Buyya, R.}, \bibinfo{author}{Winton, L.}, \bibinfo{year}{2004}.
\newblock \bibinfo{title}{A grid service broker for scheduling distributed data-oriented applications on global grids}, in: \bibinfo{booktitle}{Proceedings of the 2nd Workshop on Middleware for Grid Computing}, \bibinfo{publisher}{Association for Computing Machinery}, \bibinfo{address}{New York, NY, USA}. p. \bibinfo{pages}{75–80}.
\newblock \URLprefix \url{https://doi.org/10.1145/1028493.1028506}, \DOIprefix\doi{10.1145/1028493.1028506}.
\bibitem[{Wei et~al.(2021)Wei, Rahman, Cheng and Wang}]{wei2021joint2}
\bibinfo{author}{Wei, X.}, \bibinfo{author}{Rahman, A.}, \bibinfo{author}{Cheng, D.}, \bibinfo{author}{Wang, Y.}, \bibinfo{year}{2021}.
\newblock \bibinfo{title}{Joint optimization across timescales: Resource placement and task dispatching in edge clouds}.
\newblock \bibinfo{journal}{IEEE Transactions on Cloud Computing} \bibinfo{volume}{PP}, \bibinfo{pages}{1--1}.
\newblock \DOIprefix\doi{10.1109/TCC.2021.3113605}.
\bibitem[{Wei and Wang(2021)}]{wei2021joint1}
\bibinfo{author}{Wei, X.}, \bibinfo{author}{Wang, Y.}, \bibinfo{year}{2021}.
\newblock \bibinfo{title}{Joint resource placement and task dispatching in mobile edge computing across timescales}, in: \bibinfo{booktitle}{2021 IEEE/ACM 29th International Symposium on Quality of Service (IWQOS)}, pp. \bibinfo{pages}{1--6}.
\newblock \DOIprefix\doi{10.1109/IWQOS52092.2021.9521283}.
\bibitem[{Wright(2015)}]{Wright2015CoordinateDA}
\bibinfo{author}{Wright, S.J.}, \bibinfo{year}{2015}.
\newblock \bibinfo{title}{Coordinate descent algorithms}.
\newblock \bibinfo{journal}{Mathematical Programming} \bibinfo{volume}{151}, \bibinfo{pages}{3 -- 34}.
\newblock \URLprefix \url{https://api.semanticscholar.org/CorpusID:15284973}.
\bibitem[{Xhafa and Abraham(2010)}]{Xhafa2010Computational}
\bibinfo{author}{Xhafa, F.}, \bibinfo{author}{Abraham, A.}, \bibinfo{year}{2010}.
\newblock \bibinfo{title}{Computational models and heuristic methods for grid scheduling problems}.
\newblock \bibinfo{journal}{Future Generation Computer Systems} \bibinfo{volume}{26}, \bibinfo{pages}{608--621}.
\newblock \DOIprefix\doi{10.1016/j.future.2009.11.005}.
\bibitem[{Zeng et~al.(2024)Zeng, Wang, Zeng, Li, Shi and Huang}]{ZENG2024121}
\bibinfo{author}{Zeng, C.}, \bibinfo{author}{Wang, X.}, \bibinfo{author}{Zeng, R.}, \bibinfo{author}{Li, Y.}, \bibinfo{author}{Shi, J.}, \bibinfo{author}{Huang, M.}, \bibinfo{year}{2024}.
\newblock \bibinfo{title}{Joint optimization of multi-dimensional resource allocation and task offloading for qoe enhancement in cloud-edge-end collaboration}.
\newblock \bibinfo{journal}{Future Generation Computer Systems} \bibinfo{volume}{155}, \bibinfo{pages}{121--131}.
\newblock \URLprefix \url{https://www.sciencedirect.com/science/article/pii/S0167739X24000311}, \DOIprefix\doi{https://doi.org/10.1016/j.future.2024.01.025}.

\end{thebibliography}

\end{document}